\documentclass[11pt]{article}
\usepackage{latexsym, amsfonts}

  \newtheorem{Theorem} {Theorem} [section]
  
  \newtheorem{Proposition} [Theorem] {Proposition}
  \newtheorem{Lemma} [Theorem] {Lemma}

  \newtheorem{Corollary} [Theorem] {Corollary}
  \newtheorem{rem} [Theorem] {}              
  \newtheorem{example} [Theorem] {Example}
  
  \newenvironment{Example}
  {\begin{example} \normalfont}
  {\end{example}}
  \newenvironment{Remark}[1][Remark]
  {
    \begin{rem} \normalfont
      {\bf #1:}}
  {  \end{rem} }

  \newenvironment{Proof}[1][Proof] 
  {       
    \emph{#1:} 
    \setlength{\parskip}{0ex plus0ex minus0ex}}
  { \hspace*{\fill} \ensuremath{\Box}   }

    \newcommand {\ldd}{:\!\!}
    \newcommand {\rdd}{\!\!:}
    \newcommand {\wick}[1]{\,\ldd{#1}\rdd\,}
    
    \newcommand {\betrag} [1] {\ensuremath{ \left\vert  #1  \right\vert } } 
    \newcommand {\norm} [2] [] {\ensuremath{ \left\Vert  #2  \right\Vert_{#1} } } 
    \newcommand {\C} {\mathbb{C}}
    \newcommand {\R} {\ensuremath{\mathbb{R}}}
    
    \newcommand {\N} {\ensuremath{\mathbb{N}}}
    \newcommand {\B} {\ensuremath{\mathcal{B}}}
    \newcommand {\F}  {\ensuremath {\mathcal{F}}}
    \renewcommand {\S} {\mathcal{S}}
    
    \newcommand {\om} {\omega}

    \newcommand {\skalprod} [3] [] {\ensuremath{ \left\langle #2,#3 \right\rangle_{#1}}}
    \newcommand {\spec} {\mathrm{spec}}
    
    \newcommand {\W} {\mathcal{W}}

    \newcommand {\G}  {\ensuremath {\mathcal{G}}}
    \newcommand {\g}  {\ensuremath {\mathsf{G}}}
    \newcommand {\cN} {\mathcal{N}}
    \newcommand {\sN} {\mathsf{N}}
    \renewcommand {\P} {\mathcal{P}}
    \newcommand {\sP} {\mathsf{P}}
    \newcommand {\Veff} {V_{\mathrm{eff}}}
    \newcommand {\eps} {\varepsilon}
    \newcommand {\K} {\mathbf{K}}
    \newcommand {\mr} [1] {\mathrm{#1}}
    \newcommand {\E} {\mathbb{E}}
    
    \makeatletter
    \@addtoreset{equation}{section}
    \makeatother

\begin{document} 
\title
{\Large \bf Ground state properties of the Nelson Hamiltonian - \\A Gibbs 
measure-based approach}
\author{
\small Volker Betz, Fumio Hiroshima{\thanks{Department of Mathematics and Physics, Setsunan University, 
Osaka 572-8508}}, J\'ozsef L\H{o}rinczi, Robert A. Minlos {\thanks{Dobrushin Institute for 
Information Transmission Problems of the Russian Academy of Sciences,
Bol'shoy Karetny per. 19, Moscow, 101447}} and Herbert Spohn\\[0.1cm]
{\it \small Zentrum Mathematik, Technische Universit\"at M\"unchen} \\
{\it \small Gabelsbergerstr. 49, 80290 M\"unchen, Germany} \\}

\date{}
\maketitle

\date
\begin{abstract}
The Nelson model describes a quantum particle coupled to a scalar Bose field. We study properties of its ground state 
through functional integration techniques in case the particle is confined by an external potential.
We obtain bounds on the average and the variance of the Bose field both in 
position and momentum space, on the distribution of the number of bosons, and on the position space distribution of the particle.
\end{abstract}
\vspace{0.5cm}

\section{Introduction} \label{Section 1}
Ground states in quantum mechanics can be analysed through two in essence distinct techniques. The obvious choice is the 
eigenvalue equation, $H \psi = E \psi$, which after all serves as a definition of the ground state. The second route is more indirect
and uses positivity properties of the semigroup $e^{-tH}, t \geq 0$, which happen to be valid for many models. Through a Trotter 
type formula one can then represent ground state expectation values as averages with respect to a certain probability measure 
on a function space. This measure has the structure of a Gibbs measure and methods from statistical mechanics become available.

The standard folklore is that for systems with a few degrees of freedom the eigenvalue equation is the more powerful approach, whereas
for quantum fields with an infinite number of degrees of freedom one should employ functional integration. In fact in the latter case,
the Hamiltonian $H$ is in general not available as a self-adjoint operator on some Hilbert space, and one uses functional integration
techniques to define $H$ in the first place. 

In this paper we investigate the Nelson model of a quantum particle confined by an external potential and coupled to a scalar Bose field. This 
is a borderline case: the model has a well-defined Hamiltonian $H$, cf. Section \ref{Section 2} below, as well as  a natural functional 
measure. There has been growing interest in this model recently in connection with a rigorous control of resonances and radiation damping
 \cite{BFS..}. Here we take up the technique of functional integration with the goal of establishing bounds on ground state expectations of 
physical interest.

A basic qualitative question is how the coupling to the field modifies the localization of the particle. We will prove a pointwise
exponential bound, even a superexponential bound if the potential is sufficiently confining, for the ground state density of the particle.
They support the physical picture that coupling enhances localization. For the Bose field we study the fluctuations, which turn out to
increase through the presence of the particle, and the average density in position and momentum space. For the latter we prove 
upper and lower bounds which are sharp enough to pin down the infrared divergent behaviour. Finally we establish superexponential bounds on the 
distribution of the boson number.

The core of our paper is a ``dictionary'' which translates ground state expectations in Fock space into averages over the Gibbs measure. 
When this translation is applied to quantities of physical interest, the aforementioned
bounds turn out to be a consequence of elementary inequalities. The only extra tool that we need is a  diamagnetic type inequality for estimating
the position density of the particle.
Some of our bounds, possibly in weaker form, have been proved before by other
means; we refer to Section 6 for a discussion.

The Nelson model has the special feature that, as first observed by Feynman \cite{FH65}, one can integrate over the field degrees of freedom 
resulting in an effective action for the particle. Nelson \cite{Ne64}
  used this method in a study of the ultraviolet limit, which turned out to be
the gateway to his famous work on Markov random fields. Since then the understanding of the probabilistic structure of the functional measure 
corresponding to Nelson's model has improved considerably; we use the occasion to provide a concise and self-contained framework in Sections
\ref{Section 3} and \ref{Section 4}. 

The Nelson model with massless bosons is both infrared and ultraviolet divergent. As proved by Nelson \cite{Ne64a} through operator techniques, the latter 
is of a rather mild nature, since only the energy has to be renormalized. In this work we simply assume the appropriate cutoffs at
small and large $k$ to hold so that the Hamiltonian $H$ of (\ref{Ham}) is a self-adjoint operator in Fock space with a unique ground state.
The functional integral for the Nelson model with massless bosons in dimension $d \geq 3$ is studied in \cite{LMS00}.  
The construction of the appropriate functional measure relies
on a cluster expansion for the effective Gibbs measure on particle trajectories \cite{LM00}. This model is infrared divergent 
in $d=3$ and convergent for $d>3$. Infrared divergence means in the language of functional integration 
that the time $t=0$ path measure is singular with respect to the free $t=0$ measure.
In fact, this measure is absolutely continuous with respect to an appropriately shifted Gaussian measure, which then leads to a 
renormalized Hamiltonian $H_{\mr{ren}}$ in Fock space \cite{LMS01}. Arai \cite{Ar00} studies  $H_{\mr{ren}}$ through operator techniques.

We also refer to the monumental work of J. Fr\"ohlich \cite{Fr73,Fr74} where ground state properties of the Nelson model with
zero external potential are studied, including the removal of ultraviolet and infrared cutoffs. 

\section{Representation in Fock space} \label{Section 2}

The Hamiltonian of the model in Fock space is the operator 
\begin{equation} \label{Ham}
H := H_{\mr{p}} \otimes 1 + 1 \otimes H_{\mr{f}} + H_{\mr{I}}
\end{equation}
in $L^2(\R^d) \otimes \F$. We use $L^2(\R^d)$ to denote the set of the square integrable functions on with respect to Lebesgue measure on $\R^d$, 
while we will write $L^2(\mu)$ for the square integrable
functions with respect to any other measure $\mu$. $\F$ denotes the symmetric Fock space over $\R^d$, and
\begin{eqnarray*}
H_{\mr{p}} & = & -\frac{1}{2} \Delta + V, \\
H_{\mr{f}} & = & \int \omega(k) a^{\ast}_ka_k \, dk, \\
H_{\mr{I}} & = &  \int \frac{1}{\sqrt{2 \om(k)}} \left( \hat{\varrho}(k)e^{ikq}a_k + \hat{\varrho}(-k)e^{-ikq}a_k^{\ast}\right) \, dk.
\end{eqnarray*}

We require the potential $V: \R^d \to \R$ of the Schr\"odinger operator $H_{\mr{p}}$ to be of the form $V^+ - V^-$ with $V^+, V^- > 0$, $V^-$ in the Kato class
$K_d$ (see \cite{Si82}) and $V^+$ locally in $K_d$. In particular, $V$ can be the sum of a continuous function that is bounded below and a function having
Coulomb singularities. In addition, we assume that $V$ is chosen such that $H_{\mr{p}}$ has a unique ground state, i.e. $\inf\spec(H_{\mr{p}})$ is an eigenvalue of 
multiplicity one. As for $H_{\mr{f}}$ and $H_{\mr{I}}$, we require
\begin{eqnarray}
 & & \om(k) = \bar{\om}(k) = \om(-k), \quad \varrho(k) = \bar{\varrho}(k), \label{gc1}\\
 & &  0  <  \om(k) \quad \mbox{except on a set of Lebesgue measure zero,} \label{gc2} \\
 & & \frac{\hat{\varrho}}{\sqrt{\om}} \in L^2(\R^d), \quad \frac{\hat{\varrho}}{\om} \in L^2(\R^d). \label{gc3}
\end{eqnarray}
Here and henceforth $\hat{f}$ denotes Fourier transform and $\bar{f}$ denotes the complex conjugation of $f$, and
${f}^{\vee}$ will be used for the inverse Fourier transform of $f$.

For the convenience of the reader, we briefly recall the notions concerning symmetric Fock space involved in the above formulas. 
Denote by $L^2(\R^d)^{\hat{\otimes} n}$ the space of $L^2(\R^{dn})$-functions $f$ that are symmetric in the sense that for each $k_1, \ldots, k_n \in \R^d$
and each permutation $\pi$ of $\{1, \ldots , n \}$, we have  
$f(k_1, \ldots, k_n) = f(k_{\pi(1)}, \ldots, k_{\pi(n)})$. 
The symmetric Fock space $\F$ is the set of all $F=(f_0,f_1,\ldots)  \in\bigoplus_{n=0}^{\infty} L^2(\R^d)^{\hat{\otimes}n} $ for 
which the direct sum norm,
$$ \norm[\F]{F} = \sum_{n=0}^{\infty}\norm[L^2(\R^d)^{\hat{\otimes}n}]{f_n}, $$
converges. Putting $\F^{(n)} = L^2(\R^d)^{\hat{\otimes}n}$, it follows from the polarisation formula for multilinear maps (see \cite{Ob94})
 that $\F^{(n)}$ is spanned by linear combinations of functions of the form $f^{\otimes n} = f^{\hat{\otimes}n}$ with $f \in L^2(\R^d)$ 
(where we use the convention $f^{\otimes 0} \in \C$). Thus for defining linear
operators on $\F$ it is sufficient to specify their action on these elements.  Given such an $f^{\otimes n} \in \F^{(n)}$ and $g \in L^2(\R^d)$, we define
\begin{eqnarray*}
a^{\ast}(g) f^{\otimes n} \equiv \left( \int a^{\ast}_k g(k) \, dk \right) f^{\otimes n}  & = & \sqrt{n+1} f^{\otimes n} \hat{\otimes} g \in \F^{(n+1)}, \\
a(g) f^{\otimes n} \equiv \left( \int a_k g(k) \, dk \right) f^{\otimes n}  & = & \sqrt{n} \skalprod[L^2(\R^d)]{\bar{g}}{f} f^{\otimes (n-1)} \in \F^{(n-1)} \mbox{ for } n > 0,
\end{eqnarray*} 
and $(a(g)) (\F^{(0)}) = 0$. Here, $f^{\otimes n} \hat{\otimes} g$ is given by  
$$(f^{\otimes n} \hat{\otimes} g) (k_1, \ldots, k_{n+1}) = \frac{1}{n+1} \sum_{i=1}^{n+1} \left( \prod_{j \neq i}^{n+1} f(k_j) \right) g(k_i).$$
$a^{\ast}$ is called the creation operator and $a$ the annihilation operator. Both of them are defined on the common 
domain $\{(f_0,f_1,\ldots) \in \F: \sum_{n=0}^{\infty} n \norm[L^2(\R^d)^{\hat{\otimes}n}]{f_n} < \infty \}$. Furthermore, 
$\skalprod[\F]{F}{a(g)G} = \skalprod[\F]{a^{\ast}(\bar{g})F}{G}$ with $F,G$ in the above domain.

The operator $ \int \omega(k) a^{\ast}_k a_k \, dk$ is the differential second quantisation of the multiplication operator $f \mapsto \om f$ in $L^2(\R^d)$.
In general, given an operator $B$ in $L^2(\R^d)$, the second quantisation $\Gamma(B)$ of $B$ is the operator in $\F$ with 
$$\Gamma(B) f^{\otimes n} = (Bf)^{\otimes n}.$$
If $(B_t)_{t \geq 0}$ is a contraction semigroup on $L^2(\R^d)$ with generator $A$, then it is easy to see that 
$(\Gamma(B_t))_{t \geq 0}$ is a contraction semigroup on $\F$. The generator of this semigroup is then called the differential second quantisation
of $A$. Explicitly,
\begin{equation} \label{diffsecquant}
d\Gamma(A) f^{\otimes{n}} = \sum_{i=1}^n (Af) \hat{\otimes} f^{\otimes (n-1)}
\end{equation}
for all $f \in D(A)$. It follows that
\begin{equation} \label{Fieldop}
d\Gamma(\om) f^{\otimes{n}} \equiv \left( \int \om(k) a^{\ast}_k a_k \, dk \right) f^{\otimes n} = \sum_{i=1}^n (\om f) \hat{\otimes} f^{\otimes (n-1)}.
\end{equation}
For a self-adjoint operator $A$, both $\Gamma(A)$ and $d\Gamma(A)$ are self-adjoint. 

For every $\varepsilon > 0$ there exists $b>0$ such that 
\begin{equation} \label{InfBd}
\norm{H_{\mr{I}}g} \leq \varepsilon \norm[L^2]{\frac{\hat{\varrho}}{\om}}\norm{H_{\mr{f}} g} + b \norm[L^2]{\frac{\hat{\varrho}}{\sqrt{\om}}}\norm{g}
\end{equation}
for all $g \in D(H_{\mathrm{f}})$.
Thus by the Kato-Rellich theorem, conditions (\ref{gc3}) ensure that $H$ is self-adjoint on $D(\Delta \otimes 1) \cap D(1 \otimes H_{\mathrm{f}})$ 
and bounded from below.

\section{Representation in function space} \label{Section 3}

In this section, we develop the Schr\"odinger representation of $H$, i. e. we find an operator $\tilde{H}$ which is unitarily equivalent 
to $H$ and acts in an $L^2$-space. Moreover, $\tilde{H}$ will be the generator of a Markov process. 

We begin our construction by applying the so-called ground state transformation to $H_{\mr{p}}$. Let us write $\psi_0$ for the strictly 
positive, unique ground state of $H_{\mr{p}}$. The operator of multiplication with $\psi_0$ will be denoted by $\psi_0$ as well. $\psi_0$
is a unitary map from $L^2(\psi_0^2 \, dq)$ to $L^2(\R^d)$, and thus the  ground state transform
$\tilde{H}_{\mr{p}} = \psi_0^{-1} H_{\mr{p}} \psi_0$ of $H_{\mr{p}}$ acts in 
$L^2(\psi_0^2 \, dq)$ and is unitary equivalent to $H_{\mr{p}}$.  $\tilde{H}_{\mr{p}}$ 
is the generator of a stationary $\R^d$-valued  $P(\phi)_1$-process, i.e. the stationary solution of the SDE
$$ dq_t = (\nabla(\log \psi_0))(q_t) \, dt + dB_t.$$
We will denote the path measure of the $P(\phi)_1$-process by $\cN^0$, and its stationary measure by $\sN^0$. Note that
$d\sN^0(q) = \psi_0^2(q) \, dq$.

In order to construct an $L^2$-space for the bosonic field, consider the space  $\S' = \S'(\R^d)$ of  tempered distributions and 
the space  $\S = \S(\R^d)$ of (real-valued) Schwartz functions. Write $\G$ for  the Gaussian measure
on paths $\xi = \{ \xi_t: t \in \R\}$ $(\xi_t \in \S')$ with mean $0$ and covariance
\begin{equation} \label{3.0}
 \G(\xi_s(f) \xi_t(g)) = \int \hat{f}(k) \overline{\hat{g}(k)} \frac{1}{2 \om(k)} e^{-\om(k) |t-s|} \, dk 
\end{equation}
for all $f \in \S$ with $\int |\hat{f}|^2(k)/\om(k) \, dk < \infty$. 
$\G$ is the measure of an $\S'$-valued Ornstein-Uhlenbeck process , i.e. a stationary Gaussian Markov process with state space contained in $\S'$.
The stationary measure of $\G$ will be denoted by $\g$. It is the Gaussian measure on $\S'$ with mean 0 and covariance obtained by setting 
$t=s$ in (\ref{3.0}).\\
To get some information about support properties and path continuity of $\G$, we define a Hilbert seminorm on $\S$ by
$$ \norm[\B_D]{f}^2 = \int \hat{f}(k) \max\{\om(k),1\} D(k,k') \max\{\om(k'),1\} \overline{\hat{f}(k')} \, dk \, dk',$$
where $D(k,k')$ is the integral kernel of $(- \Delta + |k|^2)^{-(d+1)}$. The completion of $\S$ with respect to $\norm[\B_D]{.}$ will be 
denoted by $\B_D$. For $\G$-almost all $\xi$, the map $t \mapsto \xi_t$ takes its values from $\B_D$
and is continuous with respect to the topology generated by $\B_D$ \cite{LM00}.

When working with the measure $\G$ it is convenient to introduce the 
Hilbert space $K$ obtained by completing $\S$ with respect to the (complex) scalar product
\begin{equation} \label{K}
\skalprod[K]{f}{g} = \int \hat{f}(k) \overline{\hat{g}(k)} \frac{1}{2 \om(k)} \, dk \qquad (f,g \in \S).
\end{equation}
Extending the action of $\xi$ to complex-valued functions by putting $\xi(f+ig) = \xi(f) + i \xi(g)$, we find that
$$\int \xi(f) \xi(g) \, d\g(\xi) = \skalprod[K]{f}{g} \quad \mbox{for all } f,g \in K.$$
In particular, the map $\xi \mapsto \xi(f)$ is a well-defined element of $L^2(\g)$ for each $f \in K$; we will denote it 
by $\xi(f)$.\\

The connection between the Fock space $\F$ and $L^2(\g)$ is given by the Wiener-It\^{o}-Segal isomorphism. In order to
describe this isomorphism, we need Wick polynomials. The Wick polynomial
of order $n$ with respect to $\g$ is defined recursively by
\begin{eqnarray} \label{Wick}
\wick{\xi(f)^0} & = & 1, \nonumber\\
\wick{\xi(f)} & = & \xi(f), \\
\sqrt{n} \wick{\xi(f_1) \ldots \xi(f_n)} & = & \wick{\xi(f_1) \ldots \xi(f_{n-1})} \xi(f_n) - \nonumber \\
 & & - \frac{1}{\sqrt{n-1}} \sum_{i=1}^{n-1} 
\skalprod[L^2(\g)]{\xi(f_i)}{\xi(f_n)} \wick{\prod_{j \neq i}^{n-1} \xi(f_j)}. \nonumber
\end{eqnarray}
The Wiener-It\^{o}-Segal isomorphism now is the map
\begin{equation} \label{WIS}
\theta: \F \to L^2(\g), \quad f_1 \hat{\otimes} \ldots 
\hat{\otimes} f_n \mapsto \wick{ \prod_{i=1}^n \xi(\sqrt{2\om(-i\nabla)} f_i)} = \wick{ \prod_{i=1}^n \xi((\sqrt{2\om} \hat{f_i})^{\vee})}
\end{equation}
A carefully done proof of the fact that $\theta$ is indeed an isomorphism can be found in \cite{HiKuPoSt93}, although there
 a different norm convention is used for the Fock space.\\

\begin{Remark} 
\hspace{0.1cm} The fact that the Fourier transform is part of our version of the Wiener-It\^o-Segal isomorphism is somewhat inconvenient 
and leads to aesthetically slightly unsatisfactory formulas. We could have avoided this by defining the Gaussian process $\G$ on distributions
that produce real numbers when applied to Fourier transforms of real-valued functions, and omitting the hats in (\ref{3.0}). However,
since this also does not seem to be the most natural thing to do, we decided to stick to the established convention \cite{LM00,LMS00}.
\end{Remark}

Let us now describe the images of $H_{\mr{f}}$ and $H_{\mr{I}}$ under $\theta$. 
From (\ref{WIS}) and (\ref{Fieldop}) it is easy to see that 
\begin{eqnarray} \label{2.3}
\tilde{H}_{\mathrm{f}} \wick{\xi(f_1) \ldots \xi(f_n)} & \equiv & (\theta H_{\mathrm{f}} \theta^{-1}) \wick{\xi(f_1) \ldots \xi(f_n)} \\
 & = & \sum_{i=1}^n \wick{\xi((\omega \hat{f_i})^{\vee}) \prod_{j \neq i}^n \xi(f_j)}. \nonumber 
\end{eqnarray}
Note that $\tilde{H}_\mathrm{f}$ is the generator of the process $\G$ \cite{Ne73}.

On the other hand, the unitary map $\psi_0 \otimes 1$ commutes with $H_{\mr{I}}$. Thus 
writing $\Theta = \psi_0^{-1} \otimes \theta$, we easily see from (\ref{Wick}) and (\ref{WIS}) that 
for $g \in L^2(\sN^0)$ and $f \in K$,
\begin{eqnarray} \label{HI}
 \tilde{H}_{\mathrm{I}} (g \otimes \wick{\xi(f)^n}) & \equiv & (\Theta H_{\mathrm{I}} \Theta^{-1}) (g \otimes \wick{\xi(f)^n}) \\
 & = &  ( g \otimes \wick{\xi(f)^n}) \xi(\varrho(.-q)) = (g \otimes \wick{\xi(f)^n}) \cdot (\xi \ast \varrho)(q). \nonumber
\end{eqnarray}
Of course, $\xi(\varrho(.-q))$ in the above means the map $(\xi,q) \mapsto \xi(\varrho(.-q))$. 
Extending (\ref{HI}) by linearity, we find that  $\tilde{H}_{\mathrm{I}}$ is the 
operator of multiplication with $(q,\xi) \mapsto (\xi \ast \varrho)(q)$.

In sum, we find
\begin{equation} \label{tildeH}
\tilde{H} \equiv \Theta H \Theta^{-1} = \tilde{H}_{\mr{p}} \otimes 1 + 1 \otimes \tilde{H}_{\mr{f}} + \tilde{H}_{\mr{I}}.
\end{equation}
The operator
$1 \otimes \tilde{H}_{\mathrm{f}} + \tilde{H}_{\mathrm{p}} \otimes 1$ acting in 
$L^2(\sN^0 \otimes \g) =  L^2(\sN^0) \otimes L^2(\g)$ is the generator of a stationary Markov process. We will denote the measure 
$\cN^0 \otimes \G$ corresponding to this Markov process by $\P^0$, and its stationary measure $\sN^0 \otimes \g$ by $\sP^0$.

From (\ref{tildeH}) we see that $\tilde{H}$ 
is the sum of the generator of a Markov process and a multiplication operator. Modulo technical assumptions (see below), this 
implies
\begin{equation} \label{3.4}
\skalprod[L^2(\sP^0)]{F}{e^{-t\tilde{H}}G} = \int \overline{F(q_0,\xi_0)} e^{-\int_0^t (\xi_s \ast \varrho)(q_s) \, ds} G(q_t,\xi_t) \, d\P^0, 
\qquad (F,G \in \mathcal{H}).
\end{equation}
(\ref{3.4}) is called Feynman-Kac-Nelson-formula. Nelson \cite{Ne64} proved it by explicit approximation of $\tilde{H}_{\mr{I}}$. However,
since we have path continuity of $\P^0$ and 
$H_{\mr{I}}$ is infinitesimally bounded with respect to $H_{\mr{p}} \otimes 1 + 1 \otimes H_{\mr{f}}$ (see (\ref{InfBd})) , the standard proof 
using the Trotter formula  (see e.g. \cite{Si79}) also works.

\section{Gibbs measures} \label{Section 4}
The factor $\exp(\int_0^t \xi_s \ast \varrho(q_s) \, ds ) \, d\P^0$ appearing in (\ref{3.4}) defines a finite measure on $C(\R, \R^d \times \B_D).$
Normalizing it results in a probability measure with a Gibbsian structure for finite intervals (or in ``finite volume''). 
We are going to investigate the existence  of the infinite volume limit (i.e. $t \to \infty$) of this measure.
The method we use here to prove such existence relies on the following
\bigskip\\
{\bf Main assumption: } $\tilde{H}$ has a normalized,  positive ground state $\Psi \in L^2(\sP)$.
\smallskip\\

\noindent
We will require this assumption to be fulfilled throughout the rest of the paper.

Sufficient conditions for the existence and uniqueness of an $L^2$-ground state of $\tilde{H}$ are \cite{Sp97}
\begin{itemize}
\item[(i):] $\displaystyle \hat{\varrho} / \sqrt{\om} \in L^2, \hat{\varrho}/\om \in L^2$,
\item[(ii):] $\displaystyle \hat{\varrho}/\om^{3/2} \in L^2$,
\item[(iii):] $\displaystyle \Sigma - E_\mr{p} > \int \frac{|\hat{\varrho}(k)|^2k^2}{\om(k)(2 \om(k) + k^2)} \, dk,$
\end{itemize}
where $\Sigma$ is the infimum of the essential spectrum of $H_{\mr{p}}$, and $E_\mr{p} = \inf\spec H_{\mr{p}}$. In \cite{Ge99}, more 
general particle-field couplings are allowed. When specialized to our setting, the assumptions in \cite{Ge99} correspond to $\Sigma = \infty$.

Let us briefly comment on the above conditions:\\
(i) appears in Section \ref{Section 2} as Assumption (\ref{gc3}) and was needed there to ensure existence and self-adjointness of $H$. In the context of Gibbs measures
(i) is required for the existence of the free energy $\lim_{T \to \infty} \frac{1}{T} \log(Z_T)$.\\
(ii) is called the infrared cut-off condition. Under additional assumptions on $V$ and on the coupling strength $\int |\hat{\varrho}|^2/\om \, dk$, (ii) is also 
necessary for the existence of an $L^2$ ground state \cite{LMS00}. 
Thus although we will explicitly assume (ii) to hold only in Sections \ref{Section 5} and \ref{Section 6}, implicitly it plays a role also in this section.\\
(iii) is needed for currently available proofs. For the Pauli-Fierz model with external potential, Griesemer et. al. \cite{GLL00} prove the existence of a ground state
whithout such an extra assumption. Thus one would expect (i) and (ii) to suffice. Note that 
if $\lim_{|q| \to \infty} V(q) = \infty$, then $\Sigma = \infty$ and (iii) follows from (i).

Let us write  
$$X = \{X_t:t \in \R\} = \{(q_t,\xi_t): t \in \R\}$$ 
for elements of $C(\R,\R^d \times \B_D)$, and
\begin{equation} \label{P_T}
 d\P_T(X) = \frac{1}{Z_T} \exp \left( -\int_{-T}^T (\xi_s \ast \varrho)(q_s) \, ds \right) \, d\P^0(X),
\end{equation}
for the finite volume Gibbs measure. 
Here $Z_T = \int \exp ( -\int_{-T}^T (\xi_s \ast \varrho)(q_s) \, ds ) \, d\P^0(X)$ is the partition function.

In order to state our theorem about existence of the $T \to \infty$ limit of $\P_T$, we still need some preparations. First let us 
recall the notion of local weak convergence:
For a topological space $Y$ and  
an interval $S \subset \R$, denote by $\F_S$ the $\sigma$-field over $C(\R,Y)$ generated by the point evaluations with points in $S$. A sequence
of probability measures $(\mu_n)$ on $C(\R,Y)$ is said to converge locally weakly to a measure $\mu$ if for each compact interval $S \subset \R$ and each bounded,
$\F_S$-measurable function $F$, $\lim_{n \to \infty} \mu_n(F) = \mu(F)$. 

Secondly, let $\Psi$ be the ground state of $\tilde{H}$, and put $\bar{H} = \tilde{H}-E_0$, where  $E_0 = \inf\spec(H)$ is the ground state energy of $\tilde{H}$.
Denote by $\P$ the unique probability measure on $C(\R, \R^d \times \B_D)$ characterized by the conditions 
\begin{equation} \label{3.1}
\int F \, d\P = \skalprod[L^2(\sP^0)]{\Psi}{f_1 e^{-(t_2 - t_1)\bar{H}} f_2 \ldots e^{-(t_n-t_{n-1})\bar{H}}f_n\Psi}
\end{equation}
for all $F(X) = f_1(X_{t_1}) \cdots f_n(X_{t_n})$ with
$f_1, \ldots, f_n \in L^\infty(\R^d \times \B_D)$, $t_1 < \ldots < t_n$.
Note that the r.h.s of (\ref{3.1}) in fact defines a probability measure because of $e^{-t\bar{H}} \Psi = \Psi$, $\norm[L^2(\sP^0)]{\Psi} = 1$ and Kolmogorov's
consistency theorem.

\begin{Theorem} \label{Th3.1}
$ \P_T \to \P$ in the topology of local weak convergence as $T \to \infty$.
Moreover, $\P$ fulfills the DLR-equations with respect to the family $ \{ \P_T : T > 0 \}$, i.e. for $F \in \F_{[-T,T]}$ and $\P$-almost all 
$\bar{X} \in C(\R, \R^d \times \B_D)$,
\begin{equation} \label{DLR}
  \P(F | \F_{[-T,T]^c})(\bar{X}) = \frac{1}{Z_T} \int F(X) \exp \left(-\int_{-T}^T  (\xi_s \ast \varrho)(q_s) \, ds \right) \, 
  d\P^{0,T}_{\bar{X}_{-T},\bar{X}_T}(X),
\end{equation}
where $\P^{0,T}_{\bar{X}_{-T},\bar{X}_T}(X)$ is $\P^0$ conditional on $\{ X_{\pm T} = \bar{X}_{\pm T} \}$. Hence, $\P$ is a Gibbs measure with respect
to $\P^0$ for the interaction given by $\int_{-T}^T  (\xi_s \ast \varrho)(q_s) \, ds$. 
\end{Theorem}

\begin{Proof}
Let  $S>0$ and $F \in \F_{[-S,S]}$ be bounded. Since  
$$Z_T = \skalprod[L^2(\sP^0)]{1}{e^{-2T\tilde{H}}1} = \norm[L^2(\sP^0)]{e^{-T\tilde{H}}1}^2,$$
by using the Feynman-Kac-formula and the Markov property of $\P^0$ we find that, for $T > S$,
\begin{eqnarray} \label{3.2}
\lefteqn{ \int F \, d\P_T = \frac{1}{\norm[L^2(\sP^0)]{e^{-T\tilde{H}}1}^2} \int \!\!\! \int \left(e^{-(T-S)\tilde{H}}1\right)(X_{-S})  
\left( e^{-(T-S)\tilde{H}}1\right) (X_{S}) \times }  \\
 &  &  \times \left( \int \exp \left( - \int_{-S}^S (\xi_s \ast \varrho)(q_s) \, ds \right) F(X) \, 
   d\P^{0,S}_{X_{-S},X_S}(X) \right) \, d\sP^{0}(X_{-S}) d\sP^{0}(X_S). \nonumber
\end{eqnarray}
By spectral theory, for any $\tau \in \R$ we have $e^{-(T-\tau)\bar{H}}1 \to \skalprod{\Psi}{1}\Psi$ as $T \to \infty$ in $L^2(\sP^0)$. $\Psi$ is
strictly positive, therefore $\skalprod{\Psi}{1} > 0$ and
\begin{equation} \label{3.3}
\frac{1}{\norm[L^2(\sP^0)]{e^{-T\bar{H}}1}} e^{-(T-\tau)\bar{H}} 1 \stackrel{T \to \infty}{\longrightarrow} \Psi \quad \mbox{in } L^2(\sP^0),
\mbox{ for every fixed } \tau \in \R.
\end{equation}
From this it follows that 
\begin{equation} \label{3.3a}
\frac{1}{\norm[L^2(\sP^0)]{e^{-T\tilde{H}}1}} e^{-(T-\tau)\tilde{H}} 1 \stackrel{T \to \infty}{\longrightarrow} e^{-\tau E_0} \Psi
\quad \mbox{in } L^2(\sP^0), \mbox{ for every fixed } \tau \in \R, 
\end{equation}
 and thus
\begin{eqnarray*}
 \lefteqn{ \lim_{T \to \infty} \int F \, d\P_T = \int \!\!\! \int  \Psi(X_{-S})   \Psi(X_S) e^{-2SE_0} \times}\\
 & & \times \left( \int e^{-\int_{-S}^S (\xi_s \ast \varrho)(q_s) \, ds} F(X) \, 
   d\P^{0,S}_{X_{-S},X_S}(X) \right)  \, d\sP^{0}(X_{-S}) \, d\sP^{0}(X_S) = \int F \, d\P.
\end{eqnarray*}
This shows local weak convergence, and (\ref{3.1}) as well as (\ref{DLR}) now follow from the last equation by using the Feynman-Kac formula and the
Markov property of $\P^0$.
\end{Proof}

From (\ref{3.1}) it is immediate that $\P$ is the measure of a stationary Markov process and 
\begin{equation} \label{statdens}
\frac{d \sP}{d \sP^0} = \Psi^2,
\end{equation}
where $\sP$ is the stationary measure of $\P$.

The measure $\P$ has some nice additional structure, 
which we now want to describe. Fix $\bar{q} \in C(\R, \R^d)$ and denote by $\P_T^{\bar{q}}$ the 
measure $\P_T$ conditional on $\{q = \bar{q}\}$. Note that the condition $\bar{q}$ appears as an 
upper index here, as opposed to the lower indexes used in Theorem \ref{Th3.1}. The convention we use throughout 
is that conditioning on a path is denoted by upper indices, while conditioning on points is denoted by lower indices.

$\P_T^{\bar{q}}$ is a Gaussian measure on $C(\R,\B_D)$ with mean
$$ \int \xi_t(f) \, d\P_T^{\bar{q}}(\xi) = M_{t,\bar{q}}^T(f) 
  =  - \int_{-T}^T ds \int dk \, \frac{\overline{\hat{f}(k)} \hat{\varrho}(k) e^{ik\bar{q}_s}}{2\om(k)} e^{-\om(k)|t-s|} $$
($f \in K$, $f$ real-valued, $t \in \R$) and covariance equal to that of $\G$. Since 
\begin{eqnarray} \label{Isfinite}
\lefteqn{  \int_{-T}^T ds \int dk \, \betrag{\frac{\overline{\hat{f}(k)} \hat{\varrho}(k) e^{ik\bar{q}_s}}{2\om(k)} e^{-\om(k)|t-s|}} } \nonumber\\
 & \leq & \int_{-T}^T ds \left( \int \frac{|\hat{f}|^2}{2\om} \, dk \right)^{1/2} \left( \int |\hat{\varrho}|^2  \frac{e^{-2\om|t-s|}}{2\om} \, dk \right)^{1/2} 
     \nonumber \\
 & \leq & \norm[K]{f} \left( \int \frac{|\hat{\varrho}|^2}{4\om^2} \, dk \right)^{1/2} < \infty,
\end{eqnarray}
$M_{t,\bar{q}}(f) = \lim_{T \to \infty} M_{t,\bar{q}}^{T}(f)$ exists for all $\bar{q},t$ and $f$. By the convergence theory for Gaussian measures 
it follows that $\P^{\bar{q}} = \lim_{T \to \infty} \P_{T}^{\bar{q}}$ exists in the topology of weak convergence, 
and is a Gaussian measure with mean $M_{t,\bar{q}}$ and the same covariance as $\G$.
Knowing the structure of $\P^{\bar{q}}$ for each $\bar{q}$, in order to understand $\P$ we need only study the distribution $\cN$ of $\bar{q}$ under
$\P$. This will then give us a convenient representation of $\P$ as a mixture of Gaussian measures that we will use in the next section. To obtain $\cN$, let
$f \in L^1(\cN^0)$. Then, since $\xi \mapsto \int_{-T}^T (\xi_s \ast \varrho)(q_s) \, ds$ is linear and $\G$ is a Gaussian measure,
$$ \int f(q) e^{-\int_{-T}^T (\xi_s \ast \varrho)(q_s) \, ds} \, d\P^0 = \int  f(q) \exp \left( - \int_{-T}^T \int_{-T}^T W(q_s-q_t,s-t) \, 
ds \, dt \right) d\cN^0$$
with
$$ W(q,t) = -\frac{1}{2} \int \frac{\betrag{\hat\varrho(k)}^2}{2\omega(k)} \cos(kq) \,  e^{-\omega(k) \betrag{t}} \, dk.$$
We introduce
$$ d\cN_T = \frac{1}{Z_T} \exp \left( - \int_{-T}^T \int_{-T}^T W(q_s-q_t,s-t) \, ds \, dt \right) d\cN^0, $$
and by taking $F = f \otimes 1$ in Theorem \ref{Th3.1} we have the following
\begin{Corollary} \label{Cor3.2}
There exists a probability measure $\cN$ on $C(\R, \R^d)$ such that $\cN_T \to \cN$ in the
topology of local weak convergence. $\cN$ is the measure of a stationary $\R^d$-valued process. The stationary measure of $\cN$ will be denoted 
by $\sN$. We have 
$$\int f \, d\cN = \int f\otimes 1 \, d\P,$$
for each $f \in L^1(\cN)$, and 
\begin{equation} \label{einstern}
 \int F(q,\xi) d\P(q,\xi) = \int \left( \int F(q,\xi) d\P^q(\xi) \right) d\cN(q)
\end{equation} 
for each $F \in L^1(\P)$. 
\end{Corollary}

Both $\P_T$ and $\cN_T$ are finite volume Gibbs measures relative to the reference measures $\P^0$ and $\cN^0$, respectively. 
Checking that $\cN$ admits a DLR representation (i.e. is a Gibbs measure) is, however, slightly more involved than it was for 
$\P$ because $\cN$ is no longer the measure of a Markov process. This is due to the long range pair
potential $W$ that we picked up by integration over the field.
The next theorem states that 
$\cN$ does however fulfill the DLR-equations with respect to the family of measures
\begin{equation} \label{zweistern}
 d\cN_T^{\bar{q}}(q) = \frac{1}{Z_T^{\bar{q}}} \exp \left( - {\int \!\! \int}_{\!\!\!\Lambda_T} W(q_s-q_t,s-t) \, ds \, dt \right) 
\, d \cN_{T}^{0,\bar{q}}(q),
\end{equation}
where $\cN_{T}^{0,\bar{q}}$ is the $P(\phi)_1$-measure conditioned on $ \{q(s)=\bar{q}(s) \quad \forall \betrag{s} > T \}$, and 
$$\Lambda_T = (  [-T,T] \times \R ) \cup ( \R \times [-T,T] ). $$

\begin{Theorem} \label{Th3.4}
$\cN$ is a Gibbs measure for the family $\{\cN_T^{\bar{q}}: T>0\}$.
\end{Theorem}

The proof of this theorem would interrupt the main line of the paper and is therefore moved to the Appendix. To conclude this section, 
let us note that if the infrared condition (condition (ii) above) is fulfilled, then the interaction energy between the left and right 
half-line is uniformly bounded. Thus \cite{LM00} we expect $\cN$ to be unique on the set of paths with at most logarithmic increase in this case.

\section{Ground state expectations as Gibbs averages} \label{Section 5}

We now establish an explicit formula for writing expressions of the form $\skalprod[L^2(\sP^0)]{\Psi}{B \Psi}$ as Gibbs averages 
with respect to $\cN$. Remember that  $\Psi$ is the
ground state of $\tilde{H}$. From now on we assume that 
\begin{equation} \label{IRCond}
C_{\varrho} =  2 \sup_{q \in C(\R,\R^d)} \int_{-\infty}^0 ds \int_0^{\infty} dt \,  |W(q_s-q_t,s-t)| =  \int \frac{|\hat{\varrho}|^2}{2\om^3} \, dk < \infty.
\end{equation}
Note that (\ref{IRCond}) is actually the infrared condition mentioned in the previous section. As discussed there, in general we will not have 
$\Psi \in L^2(\sP^0)$ if the infrared condition is violated, and thus it is clear that (\ref{IRCond}) will be essential for our results
below to hold.
    
For stating these results we need Wick exponentials, which for $g \in K$ are given by
\begin{equation} \label{4.0.1}
 \S' \ni \xi \mapsto \wick{\exp(\xi(g))} = \sum_{n=0}^{\infty} \frac{1}{\sqrt{n!}} \wick{\xi(g)^n}.
\end{equation}
The following formulas hold for all $f,g \in K$ and all bounded, self-adjoint operators $A$ in $L^2(\R^d)$:
\begin{eqnarray} 
\wick{\exp(\xi(g))} &  = & \exp \left( \xi(g) - \frac{1}{2} \norm[K]{g}^2 \right), \label{4.0.2} \\
\skalprod[L^2(\g)]{\wick{\exp(\xi(f))}}{\wick{\exp(\xi(g))}} & = &\exp \left( \skalprod[K]{f}{g} \right), \label{4.0.2a} \\
\tilde{\Gamma}(A) \wick{\exp(\xi(g))}& = & \wick{\exp(\xi((A\hat{g})^{\vee}))}, \label{4.0.2b} \\
\skalprod[L^2(\g)]{\wick{\exp(\xi(f))}}{d\tilde{\Gamma}(A)\wick{\exp(\xi(g))}} & = & \skalprod[K]{f}{(A\hat{g})^{\vee}} \exp  \left( \skalprod[K]{f}{g} \right). \label{4.0.2c}
\end{eqnarray}
Here we put  $\tilde{\Gamma}(A) = \theta \Gamma(A) \theta^{-1}$ and $d\tilde{\Gamma}(A) = \theta d\Gamma(A) \theta^{-1}$.
Moreover, for each  $T \in [0,\infty], q \in C(\R,\R^d)$ we define
\begin{eqnarray*}
\hat{f}_{T,q}^+(k) & = & - \int_0^T \hat{\varrho}(k) e^{ikq_s} e^{-\om(k)|s|} \, ds, \\
\hat{f}_{T,q}^-(k) & = & - \int_{-T}^0 \hat{\varrho}(k) e^{ikq_s} e^{-\om(k)|s|} \, ds,
\end{eqnarray*}
and write $f^{\pm}_q(k)$ for $f^{\pm}_{\infty,q}(k)$.
Note that 
\begin{equation} \label{f+-0}
\skalprod[K]{f_q^-}{f_q^+} = - 2 \int_{-\infty}^0 ds \int_0^{\infty} dt \, W(q_t-q_s,t-s),
\end{equation}
and 
\begin{equation} \label{f+-}
\norm[K]{f_{q}^{\pm}}^2 \leq \int \frac{\hat{\varrho}(k)^2}{2\om(k)^3} \, dk =  C_{\varrho}.
\end{equation}

\begin{Theorem} \label{Th4.1}
Let $B$ be a bounded operator on $L^2(G)$. Then
\begin{eqnarray} \label{4.1}
\skalprod[L^2(\sP^0)]{\Psi}{(1 \otimes B) \Psi} & = & \int \skalprod[L^2(\g)]{\wick{\exp(\xi(f_q^-))}}{B \wick{\exp(\xi(f_{q}^+))}} 
\times  \nonumber \\
& & \times \exp\left( 2 \int_{-\infty}^0 ds \int_0^{\infty}dt \, W(q_t-q_s,t-s) \right) \, d\cN(q). 
\end{eqnarray}
\end{Theorem}

\begin{Proof}
Put
$$ \Psi_T := \frac{1}{\norm{e^{-T\tilde{H}}1}} e^{-T\tilde{H}} 1.$$
Then by (\ref{3.4}), in $L^2$-sense we have  
\begin{equation} \label{4.2}
\Psi_T(\bar{q},\bar{\xi}) = \frac{1}{\sqrt{Z_T}} \int \exp \left( - \int_0^T (\xi_s \ast \varrho)(q_s) \, ds \right) \, d\P_{\bar{q},\bar{\xi}}^0(q,\xi),
\end{equation}
where $\P_{\bar{q},\bar{\xi}}^0 = \cN_{\bar{q}}^0 \otimes \G_{\bar{\xi}}$ denotes the measure $\P = \cN^0 \otimes \G$ conditional on 
$\{q_0 = \bar{q},\xi_0 = \bar{\xi} \}$. $\G_{\bar{\xi}}$ is a Gaussian measure with mean 
\begin{equation} \label{CondGauss1}
M_{\bar{\xi},t}(f) \equiv \int \xi_t(f) \, d\G_{\bar{\xi}}(\xi) = \bar{\xi}((e^{-|t|\om}\hat{f}) ^{\vee}) \qquad (t\in \R, f \in \S)
\end{equation}
and covariance
\begin{equation} \label{CondGauss2}
\int \xi_t(f) \xi_s(g) \, d\G_{\bar{\xi}}(\xi) - M_{\bar{\xi},t}(f) M_{\bar{\xi},s}(g) = \int \frac{\hat{f} \overline{\hat{g}}}{2\om} 
\left( e^{-\om|t-s|} - e^{-\om(|t|+|s|)} \right) \, dk, 
\quad (s,t \in \R).
\end{equation}
The proof of these formulas can be found in the Appendix.
Now the integration with respect to $\G_{\bar{\xi}}$ in (\ref{4.2}) can be carried out with the result
\begin{eqnarray} \label{4.3}
\lefteqn{ \Psi_T(\bar{q},\bar{\xi}) = \frac{1}{\sqrt{Z_T}} \int \exp(\bar{\xi}(f_{T,q}^+))  \times } \\
 & & \times \exp \left( \frac{1}{2} \int_0^T  ds \int_0^T dt \int dk \frac{|\hat{\varrho}(k)|^2}{2\om(k)}
       \cos(k(q_s-q_t)) \left( e^{-\om(k)|t-s|} - e^{-\om(k)(t+s)} \right) \right) \, d\cN_{\bar{q}}^0. \nonumber
\end{eqnarray}
By (\ref{4.0.2}) we have,
$$ \exp(\bar{\xi}(f_{T,q}^+)) = \wick{\exp(\bar{\xi}(f_{T,q}^+))} \exp \left( \frac{1}{2} \int_0^T  ds \int_0^T dt \int dk \frac{|\hat{\varrho}(k)|^2}{2\om(k)}
       \cos(k(q_s-q_t)) e^{-\om(k)(t+s)} \right),$$
and hence
\begin{equation} \label{4.4}
\Psi_T(\bar{q},\bar{\xi}) = \frac{1}{\sqrt{Z_T}} \int  \wick{\exp(\bar{\xi}(f_{T,q}^+))} \exp \left( - \int_0^T ds \int_0^T dt \, W(q_s-q_t,s-t) \right) \, 
d\cN_{\bar{q}}^0.
\end{equation}
By the time reversibility of $\cN_{\bar{q}}^0$, also 
\begin{equation} \label{4.5}
\Psi_T(\bar{q},\bar{\xi}) = \frac{1}{\sqrt{Z_T}} \int  \wick{\exp(\bar{\xi}(f_{T,q}^-))} \exp \left( - \int_{-T}^0 ds \int_{-T}^0 dt \, W(q_s-q_t,s-t) \right) \, 
d\cN_{\bar{q}}^0
\end{equation}
holds. Now we write (\ref{4.5}) for the left entry and (\ref{4.4}) for the right entry of the scalar product $\skalprod{\Psi_T}{(1 \otimes B)\Psi_T}$ and use 
the fact that for $\F_{[0,\infty[}$-measurable $f,g \in L^1(\cN^0)$,
$$\int \left( \int f \, d\cN_{\bar{q}}^0 \int g \, d\cN_{\bar{q}}^0 \right) \, d\sN^0(\bar{q}) = \int f(q^+) g(q^-) \, d\cN^0(q)$$ 
(with $q^+_s = q_s$ and $q^-_s = q_{-s}$ for  $s \geq 0$), to write $\skalprod{\Psi_T}{(1 \otimes B)\Psi_T}$ as an integral with respect to $\cN^0$.
Then we add and subtract the term $2 \int_{-T}^0 ds \int_0^T dt W(q_s-q_t,s-t)$  
in the exponent and incorporate the term with the minus sign into the measure $\cN_T$. The result reads
\begin{eqnarray} \label{4.6}
 \skalprod[L^2(\sN^0 \otimes \g)]{\Psi_T}{(1 \otimes B)\Psi_T} & = & \int \skalprod[L^2(\g)]{\wick{\exp(\xi(f_{T,q}^-))}}{B \wick{\exp(\xi(f_{T,q}^+))}} \times \\
 & & \times \exp\left( 2 \int_{-T}^0 ds \int_0^{T}dt \, W(q_t-q_s,t-s) \right) \, d\cN_T(q). \nonumber
\end{eqnarray}
This is the finite $T$ version of (\ref{4.1}). It remains to justify the passing to the limit $T \to \infty$. On the left hand side of (\ref{4.6}), this is immediate
since $\Psi_T \to \Psi$ in $L^2(\sN^0 \otimes \g)$ and $B$ is continuous. On the right hand side, we already know that $\cN_T \to \cN$ in the topology
of local weak convergence, and thus it only remains to show that the integrand converges uniformly in $q \in C(\R,\R^d)$. For the second factor of the integrand
this is a consequence of (\ref{IRCond}). As for the first factor, we find that for $k \neq 0$,
$$ | f^{\pm}_{T,q}(k)|  \leq  \frac{|\hat{\varrho}(k)|}{\om(k)} \qquad \mbox{uniformly in $T$ and $q$,}$$
and  
$$ f^{\pm}_{T,q}(k)  \stackrel{T \to \infty}{\longrightarrow}  f^{\pm}_{q}(k) \qquad \mbox{uniformly in $q$.}$$ 

Thus $\wick{\exp(\xi(f_{q}^+))}$ is well defined, and $\wick{\exp(\xi(f_{T,q}^+))} \to
\wick{\exp(\xi(f_{q}^+))}$ in $L^2(\g)$ and uniformly in $q$ by dominated convergence. 
Since the same argument applies to $f_{T,q}^-$ and $B$ is continuous, the claim follows.
\end{Proof}

Most operators of physical interest are not bounded. Therefore we need to extend formula (\ref{4.1}) to unbounded operators.

\begin{Proposition} \label{Prop4.2}
Let $B$ be a self-adjoint operator in $L^2(\g)$ with 
\begin{equation} \label{4.2.1}
\int \norm[L^2(\g)]{B \wick{\exp(\xi(f^{\pm}_{q}))}}^2 \,  d\cN(q)   < \infty.
\end{equation}
Then $\Psi \in D(1 \otimes B)$, and (\ref{4.1}) holds.
\end{Proposition}

\begin{Proof}
Let $E$ be the projection valued measure corresponding to $B$, and let $B_N = \int_{-N}^N  \lambda \, dE(\lambda)$ for $N \in \N$. Then $B_N$
is a bounded operator, hence (\ref{4.1}) holds for $B_N$. Using (\ref{IRCond}) and the Cauchy-Schwarz inequality, we have
\begin{eqnarray*}
\lefteqn{\norm[L^2(\sN^0 \otimes \g)]{(1 \otimes B_N) \Psi}^2  } \\
 & = & \int \skalprod[L^2(\g)]{\wick{\exp(\xi(f^-_{q}))}}{B_N^2 \wick{\exp(\xi(f^+_{q}))}} 
                                              e^{2 \int_{-\infty}^0 ds \int_0^{\infty} dt \, W(q_t-q_s,t-s)} d\cN(q)  \\
 & \leq & e^{C_{\varrho}} \int \norm[L^2(\g)]{B_N \wick{\exp(\xi(f^{-}_{q}))}} \norm[L^2(\g)]{B_N \wick{\exp(\xi(f^{+}_{q}))}}  
   d\cN(q)   \\
 & \leq &  e^{C_{\varrho}} \int \norm[L^2(\g)]{B \wick{\exp(\xi(f^{-}_{q}))}} \norm[L^2(\g)]{B \wick{\exp(\xi(f^{+}_{q}))}}  
   d\cN(q), 
\end{eqnarray*}
which is finite according to (\ref{4.2.1}). This shows that $\Psi \in D(1 \otimes B)$ and $(1 \otimes B_N) \Psi \to (1 \otimes B) \Psi$ as $N \to \infty$. 
On the other hand, it follows from (\ref{4.2.1}) that 
$$\wick{\exp(\xi(f^{\pm}_{q}))} \in D(B) \mbox{ for $\cN$-almost all $q$.} $$ 
From this we conclude 
$$ \skalprod[L^2(\g)]{\wick{\exp(\xi(f^-_{q}))}}{B_N \wick{\exp(\xi(f^+_{q}))}} \to 
   \skalprod[L^2(\g)]{\wick{\exp(\xi(f^-_{q}))}}{B \wick{\exp(\xi(f^+_{q}))}} $$
for $\cN$ almost all $q$ as $N \to \infty$. Since by (\ref{4.0.2a}) we have
\begin{eqnarray*} 
\lefteqn{ \skalprod[L^2(\g)]{\wick{\exp(\xi(f^-_{q}))}}{B_N \wick{\exp(\xi(f^+_{q}))}}  } \\
 & \leq & \norm[L^2(\g)]{\wick{\exp(\xi(f^-_{q}))}} \norm[L^2(\g)]{B_N \wick{\exp(\xi(f^+_{q}))}}   \\
 & \leq & e^{2 C_{\varrho}}  \norm[L^2(\g)]{B \wick{\exp(\xi(f^+_{q}))}}
\end{eqnarray*}
for all $q$, and the right hand side of the above is $\cN$-integrable, the dominated convergence theorem implies 
\begin{eqnarray*} 
\lefteqn{  \int \skalprod[L^2(\g)]{\wick{\exp(\xi(f^-_{q}))}}{B_N \wick{\exp(\xi(f^+_{q}))}} 
                                              e^{2 \int_{-\infty}^0 ds \int_0^{\infty} dt \, W(q_t-q_s,t-s)} d\cN(q) \to } \\
 & \to &   \int \skalprod[L^2(\g)]{\wick{\exp(\xi(f^-_{q}))}}{B \wick{\exp(\xi(f^+_{q}))}} 
                                              e^{2 \int_{-\infty}^0 ds \int_0^{\infty} dt \, W(q_t-q_s,t-s)} d\cN(q) 
\end{eqnarray*}
as $N \to \infty$. This finishes the proof.
\end{Proof}

We now present one minor extension and two important special cases of  (\ref{4.1}).

\begin{Corollary} \label{Cor4.2.a}
Let $g \in L^{\infty}(\R^d)$, and suppose $B$ satisfies the assumptions of Proposition \ref{Prop4.2}. Then 
\begin{eqnarray*}
\skalprod[L^2(\sP^0)]{\Psi}{ (g \otimes B) \Psi} & = & \int \skalprod[L^2(\g)]{\wick{\exp(\xi(f_{q}^-))}}{B \wick{\exp(\xi(f_{q}^+))}} 
\times  \nonumber \\
& & \times g(q_0) \exp\left( 2 \int_{-\infty}^0 ds \int_0^{\infty}dt \, W(q_t-q_s,t-s) \right) \, d\cN(q). 
\end{eqnarray*}
Here $g$ is again used to denote the operator of multiplication with $g$. 
\end{Corollary}
Note that if $B$ is chosen to be the identity operator, then we
arrive at $\skalprod[L^2(\sP^0)]{\Psi}{g \Psi} = \int g(q_0) \, d\cN$, a formula that also follows from Corollary \ref{Cor3.2}.

\begin{Corollary} \label{Cor4.2.b}
For $\beta > 0$ and $g \in K$, put
$$ M(\beta)  = \skalprod[L^2(\sP^0)]{\Psi}{e^{\beta \xi(g)} \Psi}. $$
$M$ is the moment generating function for the random variable $\xi \mapsto \xi_0(g)$ under $\P$, and 
\begin{eqnarray} \label{momgenfunct}
\lefteqn{M(\beta) =  \int e^{\beta \xi_0(g)} \, d\P(q,\xi)} \\
& = & \int \exp \left( \frac{\beta^2}{2} \int \frac{| \hat{g} |^2}{2\om} \, dk - \beta \int_{-\infty}^{\infty} ds 
\int dk \,  \frac{ \hat{\varrho}(k) \overline{\hat{g}(k)} e^{ikq_s}}{2\om(k)} e^{-\om(k)|s|} \right) \, d\cN. \nonumber
\end{eqnarray}
By (\ref{Isfinite}), $M(\beta)$ is finite for all $\beta$, and hence
\begin{equation} \label{4.4.1}
\skalprod[L^2(\sP^0)]{\Psi}{\xi(g)^n \Psi} = \frac{d^n}{d\beta^n} M(\beta) |_{\beta = 0} \quad \mbox{for all $n \in \N$.}
\end{equation}
\end{Corollary}
Note that, although (\ref{momgenfunct}) can in principle be deduced from Proposition \ref{Prop4.2}, it can be obtained more easily by
using (\ref{einstern}), i.e. by fixing $q \in C(\R,\R^d)$ and integrating the function $\xi \mapsto e^{\xi_0(g)}$
with respect to the corresponding Gaussian measure.

The next statement deals with second quantisation and differential second quantisation of operators. 

\begin{Corollary} \label{Cor4.2.c}
Let $A$ be a bounded self-adjoint operator on $L^2(\R^d)$, and   
write $\tilde{\Gamma}(A) = \theta \Gamma(A) \theta^{-1}$ and $d\tilde{\Gamma}(A) = \theta d\Gamma(A) \theta^{-1}$. Then
$\Psi \in D(\tilde{\Gamma}(A))$, $\Psi \in D(d\tilde{\Gamma}(A))$ and
\begin{eqnarray} \label{4.2.c.1}
\skalprod[L^2(\sP^0)]{\Psi}{\tilde{\Gamma}(A) \Psi}&  = & \int \exp \left( \skalprod[K]{f^-_q}{(A\hat{f}^+_q)^{\vee}} \right) 
         e^{ 2 \int_{-\infty}^{0} ds \int_0^{\infty} dt \, W(q_s-q_t,s-t)} \, d\cN(q), \nonumber \\ 
\skalprod[L^2(\sP^0)]{\Psi}{d\tilde{\Gamma}(A) \Psi}&  = & \int \skalprod[K]{f^-_q}{(A\hat{f}^+_q)^{\vee}}  \, d\cN(q).
\end{eqnarray}
\end{Corollary}

\begin{Proof}
First note that by (\ref{f+-}) and the boundedness of $A$, $\norm[K]{(A\hat{f}^{\pm}_q)^{\vee}}$ is uniformly bounded in $q$.
On the other hand, 
$$ \norm[L^2(\g)]{\tilde{\Gamma}(A) \wick{\exp(\xi(f^{\pm}_{q}))}}^2 = \exp \left( \norm[K]{(A\hat{f}^+_{q})^{\vee}}^2 \right) $$
follows directly from (\ref{4.0.2a}) and (\ref{4.0.2b}). Furthermore,
$$\norm[L^2(\g)]{d \tilde{\Gamma}(A) \wick{\exp(\xi(f^{\pm}_{q}))}}^2  =  \left( \norm[K]{(A\hat{f}_q^{\pm})^{\vee}}^2 + 
              \skalprod[K]{(A\hat{f}_{q}^{\pm})^{\vee}}{f_{q}^{\pm}}^2 \right) e^{ \norm[K]{f_{q}^{\pm}}^2}$$
can be obtained from the definitions of differential second quantisation (\ref{diffsecquant}), 
Wick exponentials (\ref{4.0.1}) and of $d \tilde{\Gamma}(A)$ above. Thus (\ref{4.2.1}) is fulfilled, and Proposition \ref{Prop4.2}
now gives $\Psi \in D(\tilde{\Gamma}(A))$ and $\Psi \in D(d\tilde{\Gamma}(A))$. Now that this is established, formulas
(\ref{4.2.c.1}) follow directly from (\ref{4.1}) and (\ref{4.0.2a}) to (\ref{4.0.2c}).
\end{Proof}

\section{Bounds on ground state expectations} \label{Section 6}
We are now ready to apply the results of the previous section in order to investigate
some qualitative properties of the ground state of $H$. 
\begin{Example} \label{Ex4.3} {\bf Boson number distribution} \\
Let $P_n$ be the projection onto the $n$-th Fock space component (or $n$-boson sector). Then 
$\tilde{P}_n = \theta P_n \theta^{-1}$ is the projection onto the closure of the subspace spanned
by $\{ \wick{\xi(f)^n}, f \in K \} \subset L^2(\g)$. By (\ref{4.0.1}), we have
$$ \skalprod[L^2(\g)]{\wick{\exp(\xi(f^-_q))}}{\tilde{P}_n \wick{\exp(\xi(f^+_q))}} = \frac{1}{n!} \skalprod[K]{f^+_q}{f^-_q}^n, $$
and with (\ref{f+-0}) and Theorem \ref{Th4.1} we find 
\begin{eqnarray*}
p_n & \equiv & \skalprod[L^2(\sP^0)]{\Psi}{1 \otimes \tilde{P}_n \Psi} =
  \int \frac{1}{n!} \left( - 2 \int_{-\infty}^0 ds \int_0^{\infty} dt \, W(q_s-q_t,s-t) \right)^n \times \\
  & & \times \exp \left(2 \int_{-\infty}^0 ds \int_0^{\infty} dt \, W(q_s-q_t,s-t) \right) \, d\cN.
\end{eqnarray*}
$p_n$ is the probability of finding $n$ bosons in the ground state of $\tilde{H}$. Obviously, 
\begin{equation} \label{supexpbound}
 p_n \leq \frac{C_{\varrho}^n}{n!} e^{ C_{\varrho}}.
\end{equation}
Denoting by $N = d\Gamma(1)$ the number operator, the superexponential bound (\ref{supexpbound}) implies 
$$ \skalprod[L^2(\sP^0)]{\Psi}{e^{\alpha N} \Psi} < \infty \quad \mbox{ for each } \alpha > 0$$
and is useful in the context of scattering theory \cite{HS..}.\\
Let us now assume in addition that $W(q,t)<0$ for all $q$ and all $t$. This is true e.g. for the massive Nelson model with 
mass parameter $\kappa > 0$ and ultraviolet cutoff parameter $K \gg 1$, i.e. $\om(k) = \sqrt{k^2 + \kappa^2}$, 
$\hat{\varrho}(k) = 1_{\{|k| \leq K\}}$. Then there exists $D \leq C_{\varrho}$ with 
\begin{equation} \label{fockcomp}
 \frac{D^n}{n!} e^{- C_{\varrho}} \leq p_n \leq \frac{C_{\varrho}^n}{n!}.
\end{equation}
The right hand side of (\ref{fockcomp}) is again obvious, and
the left hand side follows from 
\begin{eqnarray*}
p_n & \geq & \frac{1}{n!} e^{- C_{\varrho}} \int   \left( - 2 \int_{-\infty}^0 ds \int_0^{\infty} dt \, W(q_s-q_t,s-t) \right)^n \, d\cN  \\
 & \geq &  \frac{1}{n!} e^{- C_{\varrho}} \left(-\int  2 \int_{-\infty}^0 ds \int_0^{\infty} dt \, W(q_s-q_t,s-t) \, d\cN \right)^n.
\end{eqnarray*}
$D$ is then the expectation of the double integral above.
\end{Example}

In the next two examples we will look at the mean value and variance of the random variable $\xi \mapsto \xi_0(g)$ under $\P$ for $g \in K$,
using the results of Corollary \ref{Cor4.2.b}.

\begin{Example} \label{Ex4.5} {\bf Average field strength} \\
For $n=1$, (\ref{4.4.1}) yields
\begin{eqnarray} \label{fieldstr}
\skalprod[L^2(\sP^0)]{\Psi}{\xi(g) \Psi} &  = & - \int dk \int_{-\infty}^{\infty} ds \, \frac{\hat{\varrho}(k)\overline{\hat{g}(k)}}{2\om(k)} e^{-\om(k)|s|} 
                                             \left( \int e^{ikq_s} \, d\cN(q) \right)  \nonumber \\
 &  = & - \int dk \int dq \psi_0^2(q) \lambda^2(q)  \frac{\hat{\varrho}(k)\overline{\hat{g}(k)} e^{ikq}}{\om(k)^2} ,
\end{eqnarray}
where $\lambda^2(q) = \int \Psi^2(\xi,q) \, d\g(\xi)$ is the stationary density of $\cN$ with respect to $\sN^0$, and $\psi_0^2$ is the density of $\sN^0$
with respect to Lebesgue measure.\\
Writing $\chi = \psi_0^2 \lambda^2$ for the position density of the particle, and taking $g$ to be a delta function in momentum space and in position space,
respectively, we find 
$$ \skalprod[L^2(\sP^0)]{\Psi}{\xi(k) \Psi} = - \frac{\hat{\varrho}(k) \hat{\chi}(k)}{(2\pi)^{d/2} \om^2(k)} \qquad (k \in \R^d), $$
and 
\begin{equation} \label{qstrength}
 \skalprod[L^2(\sP^0)]{\Psi}{\xi(q) \Psi} = (\chi \ast V_{\om} \ast \varrho)(q) \qquad (q \in \R^d),
\end{equation}
respectively. Here $V_{\om}$ denotes the Fourier transform of $- 1/\om^2$ and is the Coulomb potential for massless bosons, i.e. for $\om(k)=|k|$. 
(\ref{qstrength}) is the classical field generated by a particle with position distribution 
$\chi(q) \, dq$. Note that equality (\ref{fieldstr}) follows also from the equations of motion and the stationarity of $\Psi_0$.
\end{Example}

\begin{Example} \label{Ex4.6} {\bf Field fluctuations} \\
For $n=2$, (\ref{4.4.1}) becomes
$$ \skalprod[L^2(\sP^0)]{\Psi}{\xi(g)^2\Psi} = \int \frac{|\hat{g}(k)|^2}{2\om(k)} \, dk + \int \left( \int_{-\infty}^{\infty} ds \int dk 
\frac{\hat{\varrho}(k) \overline{\hat{g}(k)} e^{ikq_s} }{2\om(k)} e^{-\om(k)|s|} \right)^2 \, d\cN. $$
By using the previous result and the Cauchy-Schwarz inequality, we find that 
$$ \skalprod[L^2(\sP^0)]{\Psi}{\xi(g)^2 \Psi} - \skalprod[L^2(\sP^0)]{\Psi}{\xi(g)\Psi}^2 \geq \int \frac{|\hat{g}|^2}{2\om} \, dk = \int \xi(g)^2 \, d\g(\xi).$$
The latter term represents the fluctuations of the free field. We thus see that fluctuations increase by coupling the field to the particle.
\end{Example}

We now consider special cases of Corollary \ref{Cor4.2.c}.

\begin{Example} \label{Ex4.7} {\bf Average number of bosons at given momentum}\\
For real-valued $g \in L^{\infty}$ consider
$$ \int a^{\ast}_k a_k g(k) \, dk = d\Gamma(g). $$
By Corollary \ref{Cor4.2.c}, we have $\Psi \in D(d\tilde{\Gamma}(g))$. With $g$ chosen to be the indicator function of some set $B \subset \R^d$,  
$\skalprod[L^2(\sP^0)]{\Psi}{d\tilde{\Gamma}(g) \Psi}$
is the expected number of bosons with momentum within $B$. By (\ref{4.2.c.1}), 
\begin{equation} \label{4.7.1}
\skalprod[L^2(\sP^0)]{\Psi}{d\tilde{\Gamma}(g)\Psi} =
  \int dk \frac{|\hat{\varrho}(k)|^2}{2 \om(k)} g(k) \int_{-\infty}^0 ds \int_0^{\infty} dt \, e^{-\om(k)(t-s)} \int \cos(k(q_t-q_s)) \, d\cN 
\end{equation}
On the one hand, from $\cos(kx) \leq 1$ we get 
\begin{equation} \label{softphotons}
\skalprod[L^2(\sP^0)]{\Psi}{d\tilde{\Gamma}(g) \Psi} \leq \int \frac{|\hat{\varrho}(k)|^2}{2 \om(k)^3} g(k) \, dk.
\end{equation}
(\ref{softphotons}) is proven in \cite{BFS..} using the pullthrough formula.
On the other hand, from $1 - (k^2 x^2)/2 \leq \cos(kx)$ we get
\begin{eqnarray*}
\int \cos(k(q_t-q_s)) \, d\cN & \geq & 1 - \frac{k^2}{2} \left( \int (q_t^2 + q_s^2 - 2 q_tq_s) \, d\cN \right)  \\
 & \geq & 1 - k^2 \int q^2 \psi_0^2(q) \lambda^2(q) \, dq.
\end{eqnarray*}
The last inequality above follows from
$$ \int q_s q_t \, d\cN = \skalprod[L^2(\sP^0)]{\Psi q}{e^{-|t-s|\bar{H}} \Psi q} = \norm[L^2(\sP^0)]{e^{-(|t-s|/2)\bar{H}} q \Psi} \geq 0.$$
Writing $C = \int q^2 \psi_0^2(q) \lambda^2(q) \, dq$, we have for $g \geq 0$ that
$$ \skalprod[L^2(\sP^0)]{\Psi}{d\tilde{\Gamma}(g)\Psi}  \geq \int  \frac{|\hat{\varrho}(k)|^2}{2 \om(k)^3} (1 - Ck^2) g(k) \, dk.$$
The above results can be compactly (and somewhat formally) written as 
\begin{equation} \label{a^*_ka_k}
\frac{|\hat{\varrho}(k)|^2}{2 \om(k)^3} (1 - Ck^2) \leq \skalprod[L^2(\sN^0 \otimes \g)]{\Psi}{1 \otimes a^{\ast}_k a_k \Psi} \leq \frac{|\hat{\varrho}(k)|^2}{2 \om(k)^3}.
\end{equation}
Here, $a^{\ast}_ka_k$ denotes the formal expression $d\tilde{\Gamma}(\delta(\cdot - k))$.
The quantity in the middle of (\ref{a^*_ka_k}) is the expected number of bosons with momentum $k$. In particular, for the massless Nelson model,
one can see from the lower bound how the infrared divergence occurs. In this model, 
 $\om(k)=|k|, d=3,$ and $\hat{\varrho}(k) = 1_{\{ \kappa < |k| < K \}}$ with infrared cutoff parameter $0 < \kappa \ll 1$ and 
ultraviolet cutoff parameter  $K \gg 1$.
Letting $\kappa \to 0$, the expected number of bosons in the ground state with momenta in a neighbourhood of $0$ diverges.
\end{Example}

\begin{Example} \label{Ex4.8} {\bf Average number of bosons at given position}\\
We now consider the operator $A_{g}$ in Fock space given by 
$$ A_{g} = \int a^{\ast}_q a_q g(q) \, dq = d\Gamma(g(-i\nabla)) $$
for real-valued $g \in L^{\infty}(\R^d)$. Again $\tilde{A}_{g} = \theta A_{g} \theta^{-1}$. We have 
\begin{eqnarray*}
 \lefteqn{ \skalprod[K]{f_q^+}{ (\tilde{A}_{g} \hat{f}_q^-)^{\vee}}  = \int \hat{f}^+_q(k) ( \hat{g} \ast \hat{f}^-_q)(k) \frac{1}{2\om(k)} \, dk = } \\ 
 & = & \frac{1}{(2\pi)^{d/2}}  \int_{-\infty}^0 ds \int_0^{\infty} dt \int dk \int dk' \frac{\hat{\varrho}(k) \overline{\hat{\varrho}(k')}}{2\om(k)}
                 e^{i(kq_s-k'q_t)} \hat{g}(k-k') e^{-\om(k)|s|-\om(k')|t|}.
\end{eqnarray*}
Thus for $g \in L^1$, we find
$$ \skalprod[L^2(\sP^0)]{\Psi}{\tilde{A}_{g} \Psi} \leq \frac{1}{(2\pi)^{d/2}} \int dk \int dk' 
\frac{|\hat{\varrho}(k)||\hat{\varrho}(k')|}{2\om(k)^2 \om(k')} | \hat{g}(k-k') | \leq  \frac{1}{2(2\pi)^{d}} C_1 C_2 \norm[L^1]{g},$$
with $C_n = \int |\hat{\varrho}(k)|/\om(k)^n \, dk$ for $n=1,2$. Taking $g$ to be the indicator of some bounded set $B \subset \R^d$, 
$\skalprod[L^2(\sP^0)]{\Psi}{\tilde{A}_{g} \Psi}$ measures the expected number of bosons with position within $B$. From the above estimate we
see that this number is bounded by a multiple of  the volume of $B$. Moreover, it is interesting to note that 
this bound is insensitive to formally removing the 
infrared cutoff. On the other hand, the total number of bosons in the ground state is obtained by taking $g=1$ in this or the 
previous example, and we see from (\ref{a^*_ka_k}) that this quantity diverges when  the infrared cutoff is formally removed. 
\end{Example}

\begin{Example} \label{Ex4.8a} {\bf Localization of the particle}
\end{Example}
We conclude this section by showing  exponential decay of the Lebesgue-density of the stationary measure of $\cN$.
We will need the following property of $\tilde{H}$:
\begin{Proposition} \label{exdec1} 
(Diamagnetic inequality) For $f,g \in L^2(\sP^0)$ we have
$$ \skalprod[L^2(\sP^0)]{f}{e^{-t\tilde{H}}g} \leq e^{t \Veff} \skalprod[L^2(\sN^0)]{\norm[L^2(G)]{f}}{e^{-t\tilde{H}_{\mr{p}}} \norm[L^2(G)]{g}},$$
where 
$$ \Veff = \frac{1}{2} \int \frac{\betrag{\hat{\varrho}(k)}^2}{\omega^2(k)} \, dk < \infty, $$
and $\tilde{H}_{\mr{p}} = (1/\psi_0)H_{\mr{p}}\psi_0$.
\end{Proposition}
A proof can be found in \cite{Hi..}.

The second ingredient we need is a result due to Carmona \cite{Ca78}. 
For this result to hold, some mild additional restrictions 
on the single site potential $V$ are needed. We say that 
$V: \R^d \to \R$ is  in the {\em Carmona class} if there exists a breakup $V=V_1-V_2$, such that
\begin{eqnarray*}
 & & V_1 \in L^{d/2 + \eps}_{\mathrm{loc}} \mbox{ for some } \eps > 0, \mbox{ and } V_1 \mbox{ is bounded below,} \\
 & & V_2 \in L^p \mbox{ for some } p > \max\{1,d/2\}, \mbox{ and } V_2 \geq 0.
\end{eqnarray*}
Then from the proofs of Lemma 3.1 and Propositions 3.1 and 3.2 of \cite{Ca78} one can extract the following 
\begin{Lemma} \label{exdec3}
Let $V = V_1-V_2$ be of the Carmona class, and use $\W_{\bar{q}}$ to denote the measure of Brownian motion on $\R^d$ starting in $\bar{q}$.
\begin{itemize}
  \item[a)] Suppose there exist $\gamma > 0, m>0$ such that
    $$ V_1(q) \geq \gamma \betrag{q}^{2m}$$
    outside a compact set. Put $t(q) = \max \{ \betrag{q}^{1-m},1 \}$. Then for each $E>0$ there exist $D>0$ and $\delta >0$ such that
    $$ \forall \bar{q} \in \R^d: \quad e^{t(\bar{q})E} \int e^{-\int_0^{t(\bar{q})} V(q_s) \, ds} \, d\W_{\bar{q}}(q) 
       \, \leq \, D \exp(-\delta \betrag{\bar{q}}^{m+1}).$$
  \item[b)] Put $\alpha := \liminf_{\betrag{q} \to \infty} V(q)$, $t(q) := \beta \betrag{q}$ with $\beta > 0$. 
    Then for each $E \in \R$ with $E < \alpha$, there exist $D>0, \delta > 0$ and $\beta > 0$ such that 
    $$ \forall \bar{q} \in  \R^d: \quad  e^{t(\bar{q})E} \int e^{-\int_0^{t(\bar{q})} V(q_s) \, ds} \, d\W_{\bar{q}}(q) 
       \, \leq \,  D \exp(-\delta \betrag{\bar{q}}).$$
\end{itemize}
\end{Lemma}

Recall that  $\sN$ denotes the stationary measure of $\cN$, $\psi_0 \lambda$ equals
the square root of the Lebesgue density of $\sN$ (cf. Example \ref{Ex4.5})
and $E_0$ is the ground state energy of $\tilde{H}$. Our result now reads:

\begin{Theorem} \label{exdec4}
For any $V$ fulfilling the general conditions given in Section \ref{Section 2}, we have $\psi_0 \lambda \in L^{\infty}(\R^d)$. 
If, in addition, $V = V_1-V_2$ is of the Carmona class, then there exists a version of $\psi_0 \lambda$ (denoted by $q \mapsto \psi_0(q) \lambda(q)$)
for which the following statements hold: 
\begin{itemize}
  \item[a)] If $V$ satisfies  the assumptions of Proposition \ref{exdec3} a), then there exist $D,\delta>0$ with
    \begin{equation} \label{Carmona1}
    \forall q \in \R^d: \quad \psi_0(q) \lambda(q) \leq D \exp(-\delta \betrag{q}^{m+1}).
    \end{equation}
  \item[b)]  Put $\alpha := \liminf_{\betrag{q} \to \infty} V_1(q)$. If $\alpha - (E_0 + \Veff) > 0$, then there exist $D>0,\delta >0$
    such that
    \begin{equation} \label{Carmona2}
     \forall q \in \R^d: \quad \psi_0(q) \lambda(q) \leq D \exp(-\delta \betrag{q}).
    \end{equation} 
\end{itemize}
\end{Theorem}

\begin{Proof}
We first show  that $\psi_0 \lambda \in L^{\infty}(\R^d)$. Since
$\tilde{H}\Psi = E_0 \Psi$, for $h \in L^{\infty}(\R^d), h\geq 0$, the diamagnetic inequality implies 
\begin{eqnarray} \label{exdeceq1}
\int  h(q) \psi_0^2(q) \lambda^2(q) \, dq & = & \skalprod[L^2(\sP^0)]{h\Psi}{\Psi} = e^{tE_0} 
   \skalprod[L^2(\sP^0)]{h\Psi}{e^{-t\tilde{H}}\Psi} \nonumber \\
 & \leq & e^{t(\Veff + E_0)} \skalprod[L^2(\sN^0)]{h\lambda}{e^{-tH_{\mr{p}}}\lambda} \\
 & = & e^{t(\Veff + E_0)} \int h(q) \psi_0(q) \lambda(q) (e^{-tH_{\mr{p}}} \lambda \psi_0)(q) \, dq. \nonumber
\end{eqnarray}
Since we required $V$ to be in the Kato class, $e^{-tH_{\mr{p}}}$ takes $L^2(dq)$ into $L^{\infty}(dq)$ \cite{Si82}.
 Thus we can find $C \in \R$ with
\begin{equation} \label{exdeceq2}
 \int  h(q) \psi_0^2(q) \lambda^2(q) \, dq \leq C \int h(q) \psi(q) \lambda(q) \, dq,
\end{equation}
which implies $\psi_0 \lambda \in L^{\infty}$. Using this result in (\ref{exdeceq1}) and the Feynman-Kac formula to express the kernel of
$e^{-tH_{\mathrm{p}}}$, we get 
\begin{eqnarray} \label{exdeceq3}
\int h(q) \psi_0^2(q) \lambda^2(q) \, dq & \leq & e^{t(\Veff + E_0)}  \int  d\bar{q} \, h(\bar{q}) \psi_0(\bar{q}) 
                                                  \lambda(\bar{q}) \int e^{-\int_0^t V(q_s) \, ds} \psi_0(q_t) \lambda(q_t) \, d\W_{\bar{q}}(q) \nonumber\\
 & \leq & e^{t(\Veff + E_0)} \norm[L^{\infty}]{\psi_0 \lambda}^2  \int  d\bar{q} \, h(\bar{q}) \int e^{-\int_0^t V(q_s) \, ds} \, d\W_{\bar{q}}(q).
\end{eqnarray}
The version of $\psi_0 \lambda$ mentioned above can now be explicitly defined by
$$ \psi_0^2(q) \lambda(q)^2 = \limsup_{n \to \infty} \int h_{q,n}(x) \psi_0^2(x) \lambda^2(x) \, dx, $$
where $h_{q,n}$ is any fixed sequence of $L^1$-functions converging in $L^1$ to a delta peak at $q$. We now use this sequence in (\ref{exdeceq3}). Since 
$\int \exp(\int_0^t V(q_s) \, ds) \, d\W_{\bar{q}}$ is continuous in $\bar{q}$ and finite for all $\bar{q}$, 
the right hand side of (\ref{exdeceq3}) converges and we have 
$$ \psi_0^2(\bar{q}) \lambda^2(\bar{q}) \leq e^{t(\Veff + E_0)} \norm[L^{\infty}]{\psi_0 \lambda}^2 \int e^{-\int_0^t V(q_s) \, ds} \, d\W_{\bar{q}}(q). $$
This inequality is valid for each $t > 0$, and therefore in case $V$ is in the Carmona class, 
we can use Proposition \ref{exdec3} with $E$ replaced by $\Veff + E_0$ to conclude the proof.
\end{Proof}

A version of the preceding result already appears in \cite{BFS..}. There it is shown that 
$\psi_0(q) \lambda(q)  \exp(\alpha q) \in L^1(dq)$ for some $\alpha > 0$, 
while the present results (when applicable) imply  $\psi_0(q) \lambda(q)  \exp(\alpha q) \in L^{\infty}(dq)$ in case of a decaying external potential $V$
and superexponential localization in case of growing potentials.

\section{Appendix}

\subsection{Conditional Gaussian measures}

In the first part of this Appendix we prove formulas (\ref{CondGauss1}) and (\ref{CondGauss2}). In fact, we give a simple and powerful
method for explicitly calculating certain conditional Gaussian measures. 
This method must be known in some form, but we could not find it in the literature. \\
Complete the space $\S(\R^{d+1})$ with respect to a Hilbert seminorm $\norm[\K]{.}$ and denote its closure by $\K$.
Consider on $\S'(\R^{d+1})$ the Gaussian measure $\gamma$ with mean $0$ and covariance 
\begin{equation} \label{gencov}
 \int \eta(f) \eta(g) \, d\gamma(\eta) = \skalprod[\K]{f}{g} \quad (f,g \in \K).
\end{equation}
The $\sigma$-field for $\gamma$ is generated by $\{ \eta \mapsto \eta(f): f \in \K \}$. Consider now a closed subspace $\K_0 \subset \K$
and denote by $\F_0$ the $\sigma$-field generated by $\{ \eta \mapsto \eta(f): f \in \K_0 \}$. Moreover, write $P_0$ for the projection 
onto $\K_0$. By writing $f \in \K$ as $f = P_0f + f^{\perp}$, we find 
\begin{eqnarray} \label{Charfkt}
\E_{\gamma}\left( e^{i \eta(f)} | \F_0 \right) (\bar{\eta}) & = & \E_{\gamma}\left( e^{i \eta(P_0f)} e^{i \eta(f^{\perp})} | F_0 \right)  (\bar{\eta}) = 
          e^{i \bar{\eta}(P_0f)} \E_{\gamma} \left( e^{i \eta(f^{\perp})} | \F_0 \right) (\bar{\eta}) \nonumber \\
 & = &  e^{i \bar{\eta}(P_0f)} \E_{\gamma} \left( e^{i \eta(f^{\perp})} \right) = \exp \left(i \bar{\eta}(P_0f) - \frac{1}{2} \norm[\K]{f^{\perp}}^2 \right).
\end{eqnarray}
Equalities in the above equation are in $L^2(\gamma)$, and the third equality is due to the fact that independence with 
respect to $\gamma$ is equivalent with orthogonality in $L^2(\gamma)$. For $\bar{\eta} \in \S'(\R^{d+1})$, we denote by $\gamma_{\bar{\eta}}$ the 
Gaussian measure with mean $\bar{\eta}(P_0f)$ and covariance
$$ (f,g) \mapsto \skalprod[\K]{f^{\perp}}{g^{\perp}} = \skalprod[\K]{f}{g} - \skalprod[\K]{P_0f}{P_0g}. $$
It follows from (\ref{Charfkt}) that the map 
$$(\S'(\R^d),L^1(\gamma)) \to \R, \qquad (\bar{\eta},F) \mapsto \int F(\eta) \gamma_{\bar{\eta}}(\eta)$$
is a version of the regular conditional probability $\E_{\gamma}(.|\F_0)$. \\

To specialize to our context, we take for $\K$ the closure of $\S'(\R^{d+1})$ with respect to the norm associated with the scalar product   
$$ \skalprod[\K]{f}{g} = \int \hat{f}(k,\kappa) \hat{g}(k,\kappa) \frac{1}{\om^2(k) + \kappa^2} \, dk \, d\kappa \quad (k \in \R^d, \kappa \in \R) $$
and derive from this the Gaussian measure $\gamma$ according to (\ref{gencov}).
By performing the corresponding Fourier integration, it can be checked that a distribution of the form $f \otimes \delta_t$ (with $f \in K$, cf. (\ref{K}),  
and $\delta_t$ denoting the delta-peak at $t \in \R$) is an element of $\K$, and that the $\S'(\R^d)$-valued stochastic process 
$ \{ \xi_t(f) = \eta (f \otimes \delta_t), f \in K, t \in \R \} $ coincides in law with the process $\G$. Moreover, by taking $\K_0$ to
be the closure of the set $\{ f \otimes \delta_0 : f \in K \}$, we find that
$$ \widehat{P_0(g \otimes \delta_t)} = e^{-t \om} \widehat{g \otimes \delta_0} \quad \mbox{for } g \in K. $$
This can be used in the above general result to obtain (\ref{CondGauss1}) and (\ref{CondGauss2}). \\

\subsection{Proof of Theorem \ref{Th3.4}}

Before we prove the theorem,  by showing compatibility \cite{Ge88} we first make sure that the family of measures
$\{ \cN_T^{\bar{q}}: T > 0 \}$ (cf. (\ref{zweistern})) has a chance to fulfill DLR equations.
Remember that 
$$\Lambda_T = (  [-T,T] \times \R ) \cup ( \R \times [-T,T] ). $$

\begin{Lemma} \label{LeA.3}
The family $\{ \cN_T^{\bar{q}}: T > 0 \}$ is compatible.
\end{Lemma}

\begin{Proof}
We have to check that for $T>S$, measurable $A \subset C(\R,\R^d)$ and $\bar{q} \in C(\R,\R^d)$:
$$ \cN_T^{\bar{q}} \left( \cN_S^{\bullet}(A) \right) \equiv \int  \cN_S^{q}(A) \, d\cN_T^{\bar{q}}(q) = \cN_T^{\bar{q}}(A).$$
Here and henceforth we write $\cN_T^{\bar{q}}(f)$ instead of $\E_{\cN_T^{\bar{q}}}(f)$ in order to avoid too many subscript levels.
By a monotone class argument, we may assume $A$ to be of the form $A= A_1 \cap A_2 \cap A_3$ with $A_1 \in \mathcal{F}_{[-S,S]}$, 
$A_2 \in \mathcal{F}_{[-T,T] \setminus [-S,S]}$ and $A_3 \in \mathcal{F}_{[-T,T]^c}$. 
Writing $\mathcal{T}_T$ for $\mathcal{F}_{\R \setminus [-T,T]}$,
it is clear from the definition of $\cN_{T}^{0,\bar{q}}$ that for $S < T$ and $f \in L^1(\cN_{T}^{0,\bar{q}})$,
\begin{equation}
\cN_{T}^{0,\bar{q}} (f | \mathcal{T}_S) = \cN_{S}^{0,\bar{q}}(f)  \qquad \mbox{for } \cN_{T}^{0,\bar{q}} \mbox{-almost all } \bar{q} \in C(\R,\R^d)
\end{equation}
Plugging 
$$ f(q) = \cN_S^{q}(A_1) \exp \left( - {\int \int}_{\Lambda_T} W(q_s-q_t,s-t) \, ds \, dt \right) 1_{A_2}(q)$$ 
into this equality, and writing
$$ \mathcal{W}_{\Lambda}(q) := - {\int \!\! \int}_{\!\!\!\Lambda} W(q_s-q_t,s-t) \, ds \, dt $$
for $\Lambda \subset \R^2$, we find
\begin{eqnarray*}
\cN_T^{\bar{q}}(\cN_S^{\bullet}(A_1)1_{A_2})  & = & \cN_{T}^{0,\bar{q}}(f) = \cN_{T}^{0,\bar{q}} ( \cN_{T}^{0,\bullet}(f | \mathcal{T}_S) )  \\
 & = & \cN_{T}^{0,\bar{q}} \left( \cN_S^{\bullet}(A_1) 1_{A_2} e^{\mathcal{W}_{(\Lambda_T \setminus \Lambda_S)}} \cN_{S}^{0,\bullet}
\left( e^{\mathcal{W}_{\Lambda_S}} \right) \right)  \\
 & = &  \cN_{T}^{0,\bar{q}} \left( \cN_{S}^{0,\bullet} \left( 1_{A_1} e^{\mathcal{W}_{\Lambda_S}} \right) \frac{1}{\cN_{S}^{0,\bullet}
        \left( e^{\mathcal{W}_{\Lambda_S}} \right)}  1_{A_2} e^{\mathcal{W}_{(\Lambda_T \setminus \Lambda_S)}} \cN_{S}^{0,\bullet}
\left( e^{\mathcal{W}_{\Lambda_S}} \right) \right)  \\
 & = &  \cN_{T}^{0,\bar{q}} \left( \cN_{T}^{0,\bullet} \left( 1_{A_1} e^{\mathcal{W}_{\Lambda_S}} | \mathcal{T}_S \right) 1_{A_2}  
e^{\mathcal{W}_{(\Lambda_T \setminus \Lambda_S)}} \right) \\
 & = &  \cN_{T}^{0,\bar{q}} \left( e^{\mathcal{W}_{\Lambda_S}} e^{\mathcal{W}_{(\Lambda_T \setminus \Lambda_S)}}
1_{A_1} 1_{A_2} \right) \\
 & = & \cN_T^{\bar{q}} ( 1_{A_1 \cap A_2} ).
\end{eqnarray*}
Since furthermore $\cN_T^{\bar{q}}(A_3) = 1_{A_3}(\bar{q})$, the lemma is proven.
\end{Proof}
 
\begin{Proof} {\em (of Theorem \ref{Th3.4})}
Let $S<T$, put $\Lambda_{S,T} := ([-T,T] \times [-S,S]) \cup ([-S,S] \times [-T,T])$, and define
\begin{eqnarray*}
\W_{\Lambda_{S,T}} & := & - {\int \!\! \int}_{\!\!\! \Lambda_{S,T}} W(q_s-q_t,s-t) \, ds \, dt,  \\
d \cN^{\bar{q}}_{S,T}(q) & := & \frac{1}{Z_{S,T}^{\bar{q}}} \exp \left( - \W_{\Lambda_{S,T}}(q) \right) \, d\cN_{S}^{0,\bar{q}}.
\end{eqnarray*}
We claim that 
$$ \cN_T(\cdot | \mathcal{T}_S)(\bar{q}) = \cN_{S,T}^{\bar{q}} \quad \mbox{for $\cN_T$-almost all $\bar{q}$.}$$
To see this, note that $\cN^0(\cdot|\mathcal{T}_S)(\bar{q}) = \cN_{S}^{0,\bar{q}}(\cdot)$ and proceed exactly as in the proof of Lemma \ref{LeA.3}.
As a consequence, if $A \in \F_R$ for some $R > 0$, we have
\begin{equation} \label{Stern}
 \cN_T(A) = \cN_T(\cN_T(A|\mathcal{T}_S)) = \cN_T(\cN_{S,T}^{\bullet}(A)).
\end{equation}
As a last ingredient, we have for every $q \in C(\R,\R^d)$ and $T>S$
\begin{eqnarray*}
\betrag{\W_{\Lambda_{S,T}}(q) - \W_{\Lambda_S}(q)} & \leq &  4 \int_{-S}^S ds \int_T^{\infty} dt \int dk \, \frac{|\hat{\varrho}(k)|^2}{2\om(k)}
e^{-\om(k)|t-s|} \\
 & \leq & 8S \int e^{-\om(k)(T-S)} \frac{|\hat{\varrho}(k)|^2}{2\om^2(k)} \, dk \stackrel{T \to \infty}{\longrightarrow} 0
\end{eqnarray*}
by dominated convergence and (\ref{gc3}). Thus, 
$$ \sup_{\bar{q} \in C(\R,\R^d)} \betrag{\cN_{S,T}^{\bar{q}}(A) - \cN_S^{\bar{q}}(A)} \stackrel{T \to \infty}{\longrightarrow} 0,$$
and by taking $T \to \infty$ on both sides of (\ref{Stern}), we arrive at $\cN(A) = \cN(\cN_S^{\bullet}(A))$, which is what we wanted to show.
\end{Proof}

{\bf Acknowledgments:} R.A.M. thanks Zentrum Mathematik of Technische Universit\"at M\"unchen for warm hospitality and financial support. He also thanks the 
Russian Fundamental Research Foundation (grants 99-01-00284 and 00-01-00271), CRDF (grant NRM 1-2085) and DFG (grant 436 RUS 113/485/5) for 
financial support. J.L. thanks Schwerpunktprogramm `Interagierende stochastische Systeme von hoher Komplexit\"at' (grant SP 181/12).

\end{document}